\documentclass[prb,twocolumn,showpacs,superscriptaddress,floatfix]{revtex4-1}
\usepackage{graphicx,amsfonts,amssymb,amsmath}
\usepackage[pagebackref=true,colorlinks,linkcolor=blue,citecolor=red]{hyperref}

\def\be{\begin{equation}}
\def\ee{\end{equation}}
\def\bea{\begin{eqnarray}}
\def\eea{\end{eqnarray}}
\def\la{\langle}
\def\ra{\rangle}

 \newcommand{\Z}{\mathbb{Z}}
 
 \newcommand{\ket}[1]{|#1\rangle}

 \newcommand{\comutator}[2]{[ #1,#2 ]}

\begin{document}
\title{Quantum phase transition in the $\Z_3$ Kitaev-Potts model}

\author{Razieh Mohseninia}
\email{mohseninia@physics.sharif.ir}
\affiliation{Department of Physics, Sharif University of Technology, P.O. Box 11155-9161, Tehran, Iran.}

\author{Saeed S. Jahromi}
\email{s.jahromi@dena.kntu.ac.ir}
\affiliation{Department of Physics, K.N. Toosi University of Technology, P.O. Box 15875-4416, Tehran, Iran}

\author{Laleh Memarzadeh}
\email{memarzadeh@sharif.edu}
\affiliation{Department of Physics, Sharif University of Technology,
P.O. Box 11155-9161,
Tehran, Iran.}

\author{Vahid Karimipour}
\email{vahid@sharif.edu }
\affiliation{Department of Physics, Sharif University of Technology, P.O. Box 11155-9161, Tehran, Iran.}

\begin{abstract}
The stability of the topological order phase induced by the $Z_3$ Kitaev model, which is a candidate for fault-tolerant quantum computation,  against the local order phase induced by the 3-State Potts model  is studied. We show that the low energy sector of the Kitaev-Potts model is mapped to the Potts model in the presence of transverse magnetic field. Our study relies on  
two high-order series expansion based on continuous unitary transformations in the limits of small- and large-Potts couplings as well as mean-field 
approximation. Our analysis reveals that the topological phase of the $\Z_3$ Kitaev model breaks down to the Potts model through a first order phase transition.
We capture the phase transition by analysis of the ground state energy, one-quasiparticle gap and geometric measure of entanglement.
\end{abstract}
\maketitle

\section{Introduction}
\label{intro}
Quantum computers are much more powerful than the classical ones \cite{0,01}. However, a practical realization of such  machines is still a big challenge ahead due to the fragility of the qubits and their decoherence arising from inevitable interactions with the environment.  Over the years, several error correcting schemes have been 
developed \cite{shor,steane,gottesman} to make the proper working of a quantum computer possible, despite such interactions . Unfortunately, error correcting methods are themselves error prone due to the imperfections of the very gate which are responsible for detecting and correcting errors. Even this can be taken into account by the clever techniques of fault-tolerant quantum computation \cite{ft}.  Unfortunately the error threshold below which fault tolerant quantum computation is possible, is very low\cite{preskill}.
In order to overcome such problems, one can combine the main quantum feature of the quantum world, namely superposition of states, with the robustness of classical bits which is the result of a macroscopic number of very small entities, comprising each bit.  This has led to the idea of topological quantum computation\cite{kitaev1, kitaev2}, where qubits are formed from degenerate ground states of a topologically ordered many body system. The important property that these degenerate ground states cannot be distinguished by local measurements, protects them agains local errors arising from coupling with the environment, at least as long as temperature is sufficiently low (see below).  \\

The prototype of such topological qubits is the toric code first introduced by  Kitaev \cite{kitaev1}. Since then other models have also been developed, notably the color code of Bombin and Martin-Delgado \cite{BombinDelgado}.  In these models, information is stored in the topologically degenerate ground states
of the system and the computation is performed by braiding the quasiparticles (QP) of the model. 
Here one can use the gapped ground state and the robust nature of the topological phase of the model to protect information against local errors.
The only perturbations that cause logical error are those with length equal to the system size. It should be noted that this  applies only to zero temperatures, since it has been shown in \cite{nussinov1} and elaborated in \cite{nussinov2,chamon,horodecki} that the proliferation of topological defects is capable of destroying topological order and hence quantum information encoded in such order at any finite temperature. Of course by tuning the parameters of such models \cite{nussinov2} it is possible to increase the time scale for keeping the quantum information if the temperature is kept sufficiently low. \\

In his seminal paper, Kitaev showed that universal quantum computation is not possible with an Abelian group and in order to perform all universal gates of quantum computation, one needs to either resort
to the non-Abelian case or use other procedures like magic state distillation \cite{bravyi,earl} or add some other non-topological resources like measurements to Abelian models as in \cite{2,3,4,pachas}. 
Aside from the ability of Abelian models to perform universal quantum computation in the above sense, their robustness and stability against external perturbations is still a crucial question which has to
be investigated. Plenty of recent studies have been devoted for investigating  such questions \cite{parisa,mh,6,7,8,9,10,11,12,delgado,orus}. The motivation for such studies are not entirely based on quantum computation. In fact since topological order adds an entirely new paradigm for studying phase transition in condensed matter compared with the traditional one of symmetry breaking, it is very instructive to study various facets of this new phenomena. For example how topological order is destroyed under thermal fluctuation \cite{nussinov1, nussinov2, chamon}, how it gives way to local order in the presence of an external field like magnetic field \cite{10} are just two such broad questions. Of particular interest to us is the question of how topological order gives way to local order in the absence of external field when a local ordering interaction is added. This question was first investigated in our previous work \cite{parisa}, where we studied the competition of topological order and ferromagnetic order in a model which we called Kitaev-Ising model. There it was shown that there is a critical  coupling of Ising where topological order is replaced with ferromagnetic order. In that model we used a mean-field analysis and a simple calculation of Wilson-Loop operators to find the transition point. \\

In the context of condensed matter physics, it is important to understand the role of local degrees of freedom and their symmetry in topological order and its properties. It is mainly for such reasons, aside from the appeal of d-level states or qudits for quantum computation, that investigation of the above questions for general topological models for $d-$ level states is important. For example it is very desirable to understand the competition of $Z_d$ topological order and $Z_d$ local ferromagnetic order. The latter is a phase  possessed by the $d-$ level Potts model where the $Z_d$ symmetry is broken. The Potts model is a direct generalization of Ising model and has been extensively studied since its inception \cite{mattis}. While it is very desirable to perform such a study for general values of $d$, the analysis turns out to be very difficult. The $3-$State Potts model is an exception in the sense that the Hamiltonian is much simpler, because it has an equidistant spectrum which brings much simplification in our analysis. It is therefore natural that $Z_3$ Potts model has been studied more intensively in the series of $d-$ level Potts model, both in the classical context and in the new context in relation to topological order. \\

It has been shown that the $\Z_3$ Kitaev model is more robust against temperature than the $\Z_2$ one\cite{delgado}. The robustness of the model has also been studied in transverse magnetic field \cite{orus} and it has been shown that the topological order transforms to the magnetic order via a first order phase transition.  Here we want to extend these studies by studying the competition of $Z_3$ topological order induced by the $Z_3$ Kitaev model and the local order induced by the 3-State Potts model. We will see that the equidistant spectrum of the model leads to much technical simplification in the method we use, namely  Perturbative Continuous Unitary Transformations (PCUT)\cite{wegner,pcut1,pcut2,pcut3}.
We will show that like the Kitaev-Ising model \cite{parisa}, the Kitaev-Potts model also shows a first-order phase transition and the topological order break downs at a critical coupling of Potts. In contrast to the Kitaev-Ising model, here our study is more comprehensive in that we use three different methods for locating the transition point, namely 
we capture the phase transition by analysis of the ground state energy, one-quasiparticle gap and geometric measure of entanglement. The location of the transition turns out to be slightly different by these methods, and we provide arguments to show that with better approximations of these methods, this slight change will be removed. \\

The outline of the paper is as follows: In Sec.~\ref{Sec:kitaev} and \ref{Sec:potts}, we briefly review the $\Z_d$ Kitaev and Potts models and their essential features needed for our study. 
In Sec.~\ref{Sec:mapping},  we show that the low energy sector of the $\Z_3$ Kitaev-Potts  model can be mapped to the $3$-State Potts model 
in transverse magnetic field. 
We start our analysis of the mapped model in Sec.~\ref{Sec:method} by using mean field approximation and Perturbative Continuous Unitary Transformations method.
By applying the PCUT method to the small Potts coupling limits, we calculate the one-quasi particle (1-QP) gap and also the Geometric Measure of Entanglement. 
We further compute the ground-state energy in both small and large Potts couplings and capture the phase transition and the breakdown of the topological phase of the Kitaev model by analysis of the ground state 
energy and its derivatives, as well as the Geometric Measure of Entanglement and energy gap.
Finally, Sec.~\ref{Sec:conclude} is devoted to the conclusion.

\section{$\Z_d$ Kitaev model}\label{Sec:kitaev}
\begin{figure}[t]
\begin{center}
\includegraphics[scale=0.5]{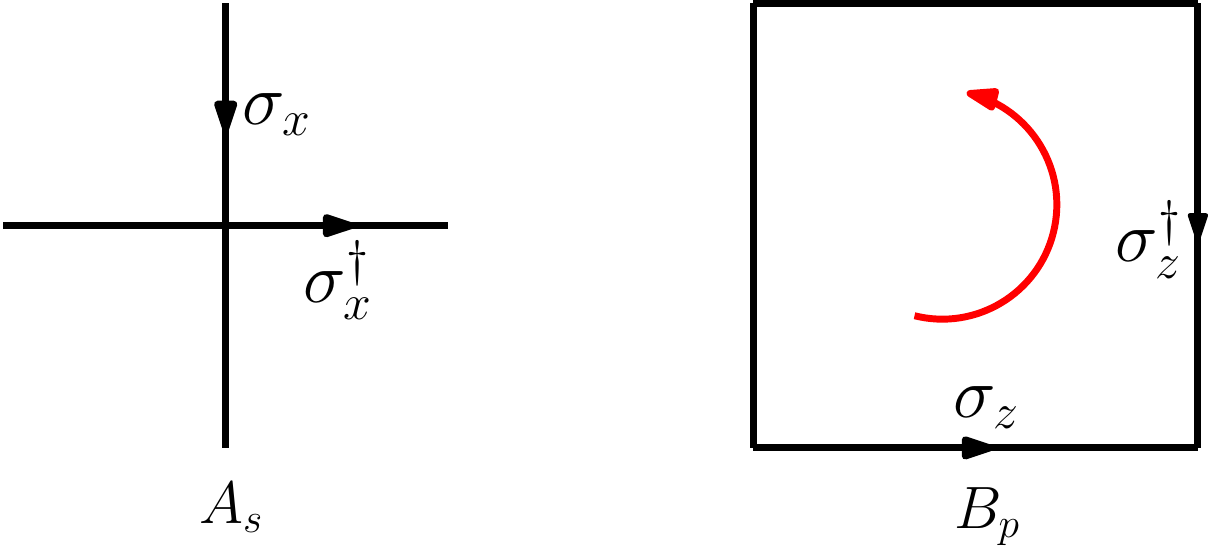}
\caption{(Color Online)The conventions for defining vertex and plaquette operators. $
A_s:= \prod_{i \in s} \sigma_{x,i}^{\pm 1}$, where the $+$ ($-$) corresponds to links with inward (outward) direction and
$
B_p:= \prod_{i \in p} \sigma_{z,i}^{\pm 1}
$,
where the $+$ ($-$) corresponds to counterclockwise (clockwise) direction of the links around plaquettes.}\label{operators}
\end{center}
\end{figure}

$\Z_d$ Kitaev model is the generalization of the Kitaev model from $\Z_2$ to $\Z_d$ group\cite{zd}. The model can be defined on any two-dimensional lattice. In the present work, we consider the model on a square lattice on which the periodic boundary conditions are imposed on both sides, where the lattice becomes  a torus. The lattice has $N$ plaquettes, $N$ vertices and $2N$ edges. The qudits live on the edges of the lattice. The Hamiltonian of the model consists of two kinds of operators, i.e. the plaquette and vertex operators. The operators are define based on the generalized Pauli operators (acting on a qudit)  as: $\sigma_x  |j\ra=|j+1 \quad $mod d$ \ra$ and $\sigma_z |j\ra= \omega^j|j\ra, \ \omega=e^{2 \pi i/d}$. These operators are not Hermitian and they don't square to $I$ except for $d$=2. They further obey the following commutation relation: $\sigma_z \sigma_x=\omega \sigma_x \sigma_z$.
The Hamiltonian of the model is given by the sum of the plaquette and vertex operators as:
\be
H_{Kitaev}:=-J \sum_s (A_s+A_s^\dagger) -K \sum_p (B_p+B_p^\dagger),
\ee 
where $s$ and $p$ denote the stars (vertices) and plaquettes respectively.  In order for the model to be exactly solvable, the 
$A_s$'s and $B_p$'s are defined such that they commute with each other. To this end  an arbitrary direction is assigned to each edge and the $A_s$ and $B_p$ operators are defined as follows:
\begin{itemize}
\item $A_s:= \prod_{i \in s} \sigma_{x,i}^{\pm 1}$, if the link's direction is inward,  $\sigma_x$ is applied, otherwise $\sigma_x^{-1}$ is,
\item $B_p:= \prod_{i \in p} \sigma_{z,i}^{\pm 1}$, by starting from a link  and moving counterclockwise,  if each link's direction is the same as moving's direction $\sigma_z$ is applied , otherwise $\sigma_z^{-1}$ is. (Fig.~\ref{operators})
\end{itemize}

These conventions lead to the commutativity of all $A_s$'s and $B_p$'s.
Let us note  that these arbitrary orientations do not have a physical significance, i.e. they lead to unitary equivalent models. That is if we change the orientations on some links (${L}$) in an arbitrary manner, we will end up with a model ($\tilde{H}$) which is iso-spectral with the original one ($H$). This unitary operator should transform $\sigma_x$ to $\sigma_x^\dagger$ and $\sigma_z$ to $\sigma_z^\dagger$ in the Hamiltonian:
\be
\tilde{H}=U H U^\dagger, \qquad \qquad U=\bigotimes_{i \in L} u_i,
\ee
where
\be 
u_i \sigma_{x,i} u_i^\dagger=\sigma_{x,i}^\dagger, \qquad \qquad u_i \sigma_{z,i} u_i^\dagger=\sigma_{z,i}^\dagger,
\ee
It is straightforward to show that $u$ is given by:
\be
u=\sum_k|d-k\rangle \langle k|.
\ee

Note that there are $2N$ stabilizers in the Hamiltonian, but only $2N-2$ of them are independent, because of the following two constraints on the torus:
 \be
\prod_sA_s=\prod_pB_p=I.
 \ee
So there are $d^2$ degenerate ground states. The ground state is the state that is stabilized by all of the star and plaquette operators simultaneously and is equal to:
\be\label{1}
|\tilde{0}\tilde{0}\ra:= \prod_s(1+A_s+A_s^2+A_s^3+ ... + A_s^{d-1}) |0\ra^{\otimes2N}.
\ee

\begin{figure}[t]
\begin{center}
\includegraphics[scale=0.5]{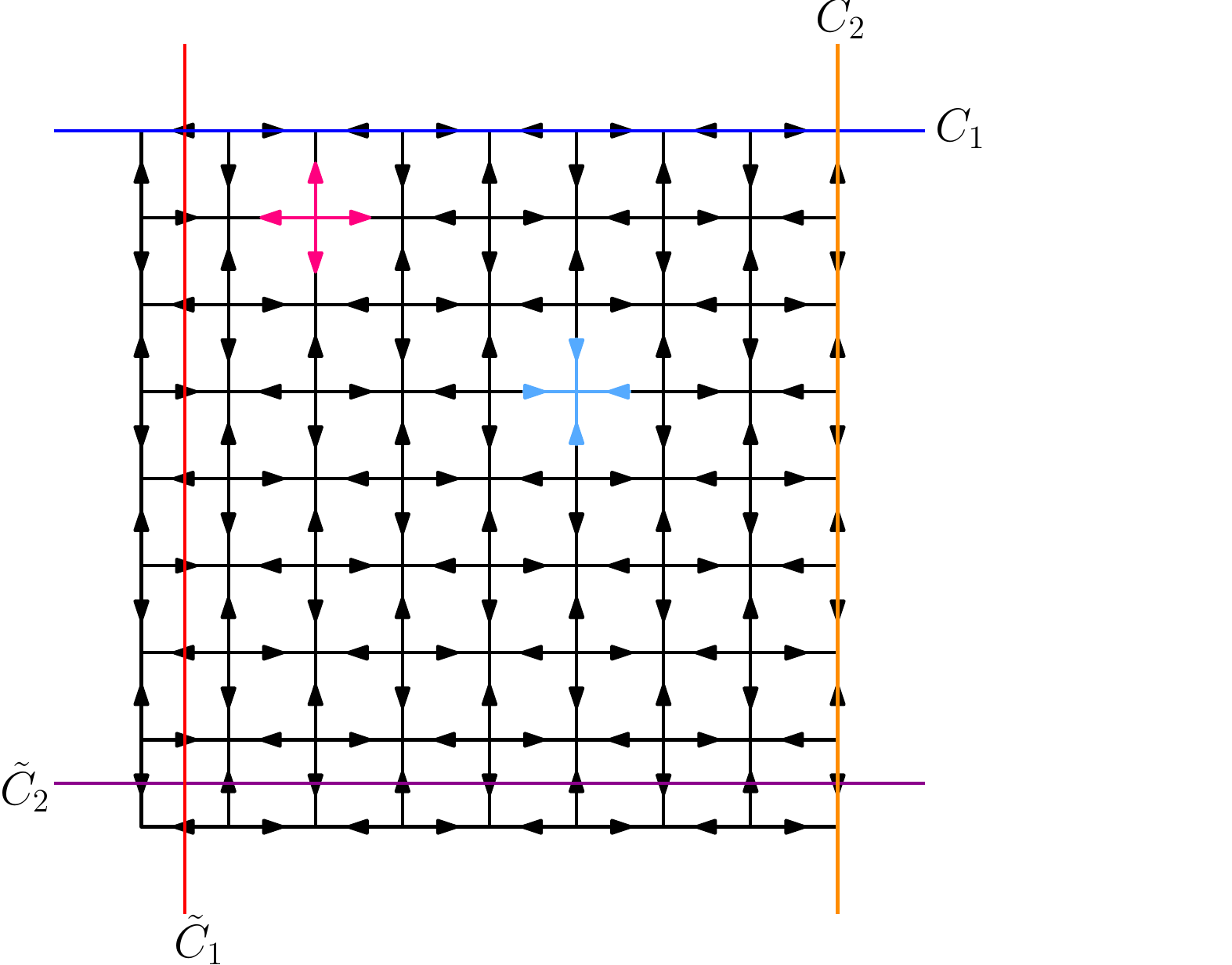}
\caption{(Color Online)Four nontrivial loops in a torus. $C_1$ and $C_2$ are in the real lattice, while $\tilde{C}_1$ and $\tilde{C}_2$ are in  the dual lattice. }\label{lattice}
\end{center}
\end{figure}

In order to construct the other $d^2-1$ degenerate ground states, we define the following four string operators:
\begin{itemize}
\item $T_{z,1}= \prod_{i \in C_1} \sigma_{z,i}^{\pm1}$, by starting from a link on the loop $C_1$ and moving on it, if each link's direction is the same as moving's direction $\sigma_z$ is applied , otherwise $\sigma_z^{-1} $ is,
\item $T_{z,2}=\prod_{i \in C_2} \sigma_{z,i}^{\pm1}$, by starting from a link on the loop $C_2$ and moving on it, if each link's direction is the same as moving's direction $\sigma_z$ is applied , otherwise $\sigma_z^{-1}$ is,
\item $T_{x,1}=\prod_{i \in \tilde{C}_1} \sigma_{x,i}^{\pm1}$, if the links's direction is the same as moving's direction on ${C}_1$, $\sigma_x$ is applied, otherwise $\sigma_x^{-1}$ is, 
\item $T_{x,2}=\prod_{i \in \tilde{C}_2} \sigma_{x,i}^{\pm1}$,  if the links's direction is the same as moving's direction on ${C}_2$, $\sigma_x$ is applied, otherwise $\sigma_x^{-1}$ is,
\end{itemize}
where $C_1$, $C_2$, $\tilde{C}_1$ and $\tilde{C}_2$ are shown in Fig.~\ref{lattice}.
All the degenerate ground states are given as:
\be
|\tilde{i}\tilde{j}\ra = T_{x,2}^j T_{x,1}^i |\tilde{0}\tilde{0}\ra, \qquad \qquad i,j=0,1,2, ... , d-1.
\ee
One can further check that:
\be
T_{z,1} |\tilde{i}\tilde{j}\ra = \omega^i |\tilde{i}\tilde{j}\ra, \qquad \qquad T_{z,2} |\tilde{i}\tilde{j}\ra =\omega^j |\tilde{i}\tilde{j}\ra.
\ee

\section{Potts model}\label{Sec:potts}
Considering any lattice of interest, the classical Potts model is defined by the following Hamiltonian \cite{mattis}:
\be
H_{Potts} = -\sum_{i,j} \delta_{s_i,s_j},\label{eq:potts}
\ee
where the sum runs over the nearest neighbor sites of the lattice ($i,j$)  and $s_i$ takes $d$ different values. One can take these values to be $d$ different roots of unity. For $d$=2, $s_i=\lbrace1,-1\rbrace$,  which reduces the Hamiltonian (\ref{eq:potts}) to the renowned Ising model. This simplification is a consequence of the following definition for the 
delta function: $\delta_{s_i,s_j}=\frac{1}{2} (1+ s_i s_j)$. Such a definition can be extended to $d$-level and the analogue formula for the Potts model reads:
\be
\delta_{s_i,s_j}= \frac{1}{2d} \sum_{r=0}^{d-1} ((s_i s_j^\ast)^r+(s_i^\ast s_j)^r),
\ee
where $s_j^\ast$ is the complex conjugate of $s_j$. For general value of $d$, the model posses $Z_d$ symmetry $s_j\longrightarrow \omega s_{j}$. 

The quantum Potts model consists of $d$-level spins (qudits). The nearest neighbor spins interact with each other by the following Hamiltonian:
\be 
H_{Potts} = - \frac{1}{2d} \sum_{\la i,j \ra} \sum_{r=0}^{d-1} \big((\sigma_{z,i} \sigma_{z,j}^\dagger)^r+(\sigma_{z,i}^\dagger \sigma_{z,j})^r\big).
\ee
The ground state of the system is the state where all  spins are polarized in the same direction. So there are $d$ degenerate ground states as follows:
\be
|\tilde{i}\ra := |i\ra^{\otimes L}, \qquad \qquad i=0,1, ... , d-1,
\ee
where $L$ is the number of spins.

\section{ The 3-State Kitaev-Potts}\label{Sec:mapping}
Our aim is to study the phase transition of the Kitaev model in presence of the Potts interaction. As our series expansion technique, i.e. the PCUT method is 
only applicable to those models with equidistant spectrum, we restrict our study to the $d=3$ or qutrits and show that $Z_3$ Kitaev and $3$-State Potts models have equidistant spectrum. The full Hamiltonian of the $\Z_3$ Kitaev model perturbed by the Potts interaction is given by:
\bea
H &=& H_{Kitaev} + \lambda H_{Potts} \nonumber \\
 \qquad  &=& -J \sum_s (A_s+A_s^\dagger) -K \sum_p (B_p+B_p^\dagger) \nonumber\\
&-&  \frac{\lambda}{6} \sum_{\la i,j \ra} \sum_{r=0}^{2} \big((\sigma_{z,i} \sigma_{z,j}^\dagger)^r+(\sigma_{z,i}^\dagger \sigma_{z,j})^r\big) . \label{eq:kitaev-potts}
\eea
where $\lambda$ is the perturbation parameter and is a measure of the strength of the Potts interaction. The perturbed Kitaev Hamiltonian is no longer exactly solvable. This is due to the fact that the $\sigma_z$ operators in the Potts model do not commute with vertex operators of the Kitaev model. However, the plaquette operators still commute with the full Hamiltonian (\ref{eq:kitaev-potts}). The ground state of the Hamiltonian (\ref{eq:kitaev-potts}) is therefore in the sector in which $B_p=1$ for all plaquette operators. In this sector the Kitaev-Potts's Hamiltonian reduces to the following form:
\begin{align}
H &=-J \sum_s (A_s+A_s^\dagger) \nonumber \\
&- \frac{\lambda}{6} \sum_{\la i,j \ra} \sum_{r=0}^{2} \big((\sigma_{z,i} \sigma_{z,j}^\dagger)^r+(\sigma_{z,i}^\dagger \sigma_{z,j})^r\big) 
-2 KN.\label{eq:K-P-simple}
\end{align}

\begin{figure}[t]
\begin{center}
\includegraphics[scale=0.5]{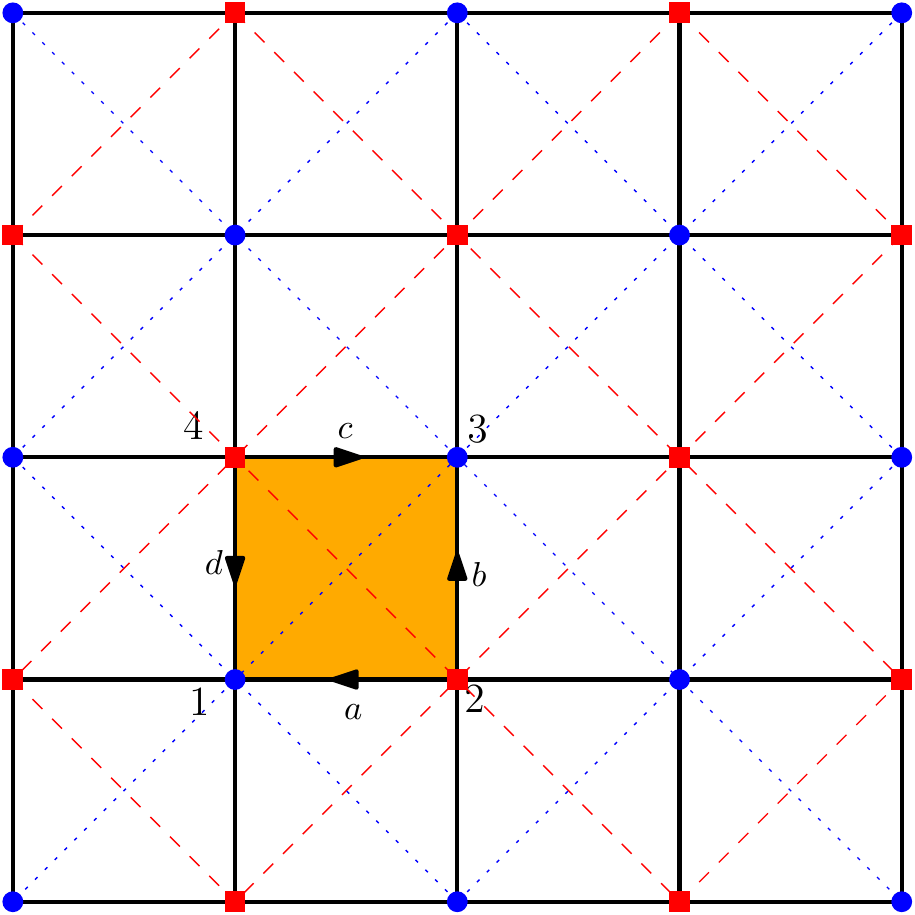}
\caption{(Color Online)In the mapped Kitaev-Potts Hamiltonian, the original lattice (the lattices with $\sqrt{N} \ast \sqrt{N}$ Plaquettes, wherein $\sqrt{N}$ is even) is de-coupled into two sublattices shown with blue dots and red dashes.}\label{mapping}
\end{center}
\end{figure} 
In order to tackle the Hamiltonian (\ref{eq:K-P-simple}), we first define the following new basis and rewrite the full Hamiltonian in this new basis:
\be
|\bold{r}\ra=|r_1,r_2,...,r_N\ra := \prod_{i} A_i^{r_i} |0\ra^{\otimes 2N},  \quad r_i=0,1,2,\label{eq:basis} 
\ee
where $i$ runs over all the vertices of the lattice. This transformation was first defined in \cite{parisa} for the the Kitaev-Ising model. 
One can readily check that $B_p |\bold{r} \ra=|\bold{r}\ra$, We can therefore interpret Eq.~(\ref{eq:basis}) as the basis for the sector in which $B_p=1$ for all $p$. Next, we determine the action of the different terms in the Hamiltonian (\ref{eq:K-P-simple}) on this new basis:
\be
A_s |\bold{r}\ra = A_s \prod_i A_i^{r_i} |0\ra^{\otimes 2N}=|r_1,r_2,...,r_s+1,...,r_N\ra,
\ee
so $A_s$ in this basis acts like generalized Pauli operator, $\sigma_x$, which we denote by $\hat{X}$. Clearly, the action of $A_s^\dagger$ can be regarded as $\hat{X}^{\dagger}$. In order to recast the Potts interaction in the new basis, we attach an arrow to each link. The direction of the arrows are illustrated in Fig.~ \ref{lattice}. The action of a Potts interaction term like $\sigma_{z,a} \sigma_{z,b}^\dagger$ in the new basis, is denoted by:
\bea
\sigma_{z,a} \sigma_{z,b}^\dagger |\bold{r}\ra &=& \sigma_{z,a} \sigma_{z,b}^\dagger A_{1}^{r_{1}} A_{2}^{r_{2}} A_{3}^{r_{3}} \prod_{i\neq 1,2,3} A_{i}^{r_i}  |0\ra^{\otimes 2N} \\ \nonumber
&=&  \omega^{r_{1}} \omega^{-r_{3}}  |\bold{r}\ra,
\eea
which means that the Potts interaction term, $\sigma_{z,a} \sigma_{z,b}^\dagger$, commutes with all of the vertex operators except $A_{1}$ and $A_{3}$, actually it acts like $\hat{Z}_{1} \hat{Z}_{3}^\dagger$ in this new basis:
\be
\sigma_{z,a} \sigma_{z,b}^\dagger \equiv  \hat{Z}_{1} \hat{Z}_{3}^\dagger.
\ee
It is straightforward to check that the following relations further hold in the new basis: 
\bea
\sigma_{z,a}^\dagger \sigma_{z,b} \equiv \hat{Z}_{1}^\dagger \hat{Z}_{3}, \nonumber \\
 \sigma_{z,c}^\dagger \sigma_{z,d} \equiv \hat{Z}_{1} \hat{Z}_{3}^\dagger, \nonumber \\
  \sigma_{z,c} \sigma_{z,d}^\dagger  \equiv \hat{Z}_{1}^\dagger \hat{Z}_{3}.
\eea
Therefore the $\Z_3$ Kitaev-Potts model in the new basis is given by:

\be
\tilde{H} =\tilde{H}_A+\tilde{H}_B - 2 K N, \label{eq:mapped-model}
\ee
where
\bea
\tilde{H}_A=-J \sum_{i \in A} (\hat{X}_i+\hat{X}^{\dagger}_i) -\frac{ \lambda} {3} \sum_{ \langle i,j \rangle\in A}\sum_{r=0} ^{2} (\hat{Z}_i \hat{Z}^{\dagger}_j)^r +(\hat{Z}_j \hat{Z}^{\dagger}_i)^r, \nonumber \\ \label{eq:HA}
\eea
and
\bea
\tilde{H}_B= -J \sum_{i \in B} (\hat{X}_i+\hat{X}^{\dagger}_i)- \frac{ \lambda} {3} \sum_{ \langle i,j \rangle\in B}\sum_{r=0} ^{2}  (\hat{Z}_i \hat{Z}^{\dagger}_j)^r + (\hat{Z}_j \hat{Z}^{\dagger}_i)^r. \nonumber \\
\eea

The Hamiltonian (\ref{eq:mapped-model}) is nothing but the sum of two Potts models in a transverse magnetic field, wherein the Potts interactions act on nearest neighbor 
vertices in the two de-coupled sublattices shown in Fig.~\ref{mapping}.

The two de-coupled Hamiltonians are exactly the same. In the following sections we present our results for $\tilde{H}_A$. The results can be extended to the 
full Hamiltonian (\ref{eq:mapped-model}), without loss of generality. From now on to the end of the paper by $H$, we mean $\tilde{H}_A$, unless stated otherwise.

\section{Methods}\label{Sec:method}
In this section, we present the solution to the mapped model by applying the Mean field approximation and Perturbative Continuous Unitary Transformations method 
to the Hamiltonian (\ref{eq:HA}).

\subsection{Mean Field approximation}\label{Sec:mean-field}
Suppose we are interested in finding the ground state energy of a given Hamiltonian. We therefore need to minimize 
$\la \Psi | H | \Psi \ra$ over all $|\Psi \ra$'s in the Hilbert space. One approximation (Mean field approximation) would be that we do the minimization only over product states. But the Hamiltonian has translational symmetry so we can search for the minimum energy only over states of the form  $|\Psi \ra= | \Phi \ra^{\otimes n}$, which have  translational symmetry. For the mapped Hamiltonian $n$ is the number of vertices in each sublattice, $n=\frac{N}{2}$.
\be
E= \min_{| \Phi \ra} {}^{\otimes n}\la \Phi |H | \Phi \ra^{\otimes n},
\ee
where $|\Phi \ra$ is a general one-qutrit state. For the mapped Hamiltonian,

\bea
H=-J \sum_{i} (\hat{X}_i+\hat{X}^{\dagger}_i) -\frac{ \lambda} {3} \sum_{ \langle i,j \rangle} \sum_{r=0} ^{2} (\hat{Z}_i \hat{Z}^{\dagger}_j)^r + (\hat{Z}_j \hat{Z}^{\dagger}_i)^r, \nonumber
\eea
we have:
\be
E_0= -J n \Big( \langle \hat{X}\rangle +\langle \hat{X}^{\dagger} \rangle \Big) - 4n \frac{ \lambda} {3}  \sum_{r=0} ^{2} \Big( \langle \hat{Z}^r \rangle \langle \hat{Z}^{-r} \rangle \Big).
\ee

Taking $| \Phi\rangle= \sum_{i=0}^{2} a_i | i\rangle$, we should minimize:
\begin{equation}
E_0= -J n \Biggl (  \sum_{i} \Bigl( a_i a_{i+1}^{\ast}+ a_{i}^{\ast} a_{i+1} \Bigl ) \Biggl) - 4n \lambda \Biggl (  \sum_{i} |{a_i}|^4 \Biggl ).
\end{equation}
\begin{figure}[t]
\begin{center}
\includegraphics[width=\columnwidth]{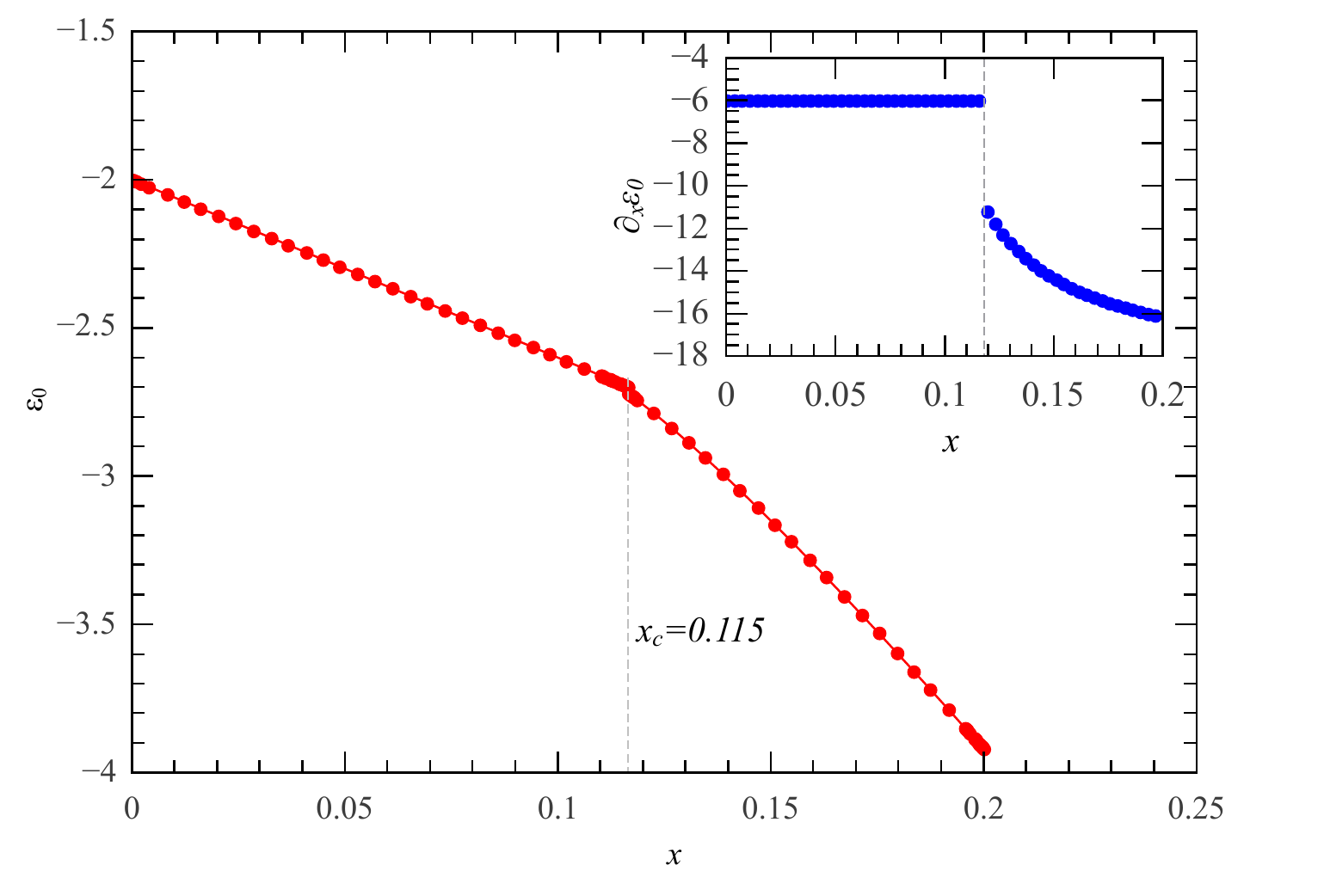}
\caption{(Color Online) Ground state energy per site, $\varepsilon_0$, as a function of x. The inset represents its first derivative. The sharp jump in the derivation of energy at $x_c\approx0.115$ signals the first order quantum phase transition.}\label{meanfield}
\end{center}
\end{figure} 
It's easy to check that the above equation recasts into:
\be
E_0= -J n  \Bigl(|a_0+a_1+a_2|^2 -1\Bigl )- 4n \lambda \Biggl ( |{a_0}|^4+|{a_1}|^4+|{a_2}|^4 \Biggl ),
\ee
that has a permutation symmetry by changing the $a_i$'s indices from $(0,1,2)$ to $(\sigma(1),\sigma(2),\sigma(3))$. If the state that minimize this expression respects the full symmetry, it should be equal to $|\Phi\ra=\frac{1}{\sqrt{3}}(|0\ra+|1\ra+|2\ra)$ and the ground state energy per site is equal to:
\be
\varepsilon_0=-2J-\frac{4}{3}\lambda.\label{eq:E-meanfield}
 \ee
Setting $J=1$, we see that Eq.~(\ref{eq:E-meanfield}) is analytic for any value of $\lambda$ and one does not observe any phase transition. However, if some of the symmetries are broken and the ground state remains invariant under only one permutation, it will be in the form of $|\Phi\ra=a_0|0\ra+a_1|1\ra+a_1|2\ra$. By setting $a_0=\sin\theta, a_1=\frac{1}{\sqrt{2}}e^{i \alpha}\cos\theta$, the ground state energy per site reduces to:
\be
\varepsilon_0= -J(\cos^2\theta+\sqrt{2}\sin2\theta \cos\alpha )- 4 \lambda( \sin^4\theta+\frac{1}{2}\cos^4\theta ).
\ee
By minimizing $\varepsilon_0$, the ground state as a function of perturbation parameter $x=\frac{2\lambda}{9J}$ (this explicit form of $x$ is chosen because it makes 
the comparison between mean-field and PCUT results easier) is given as follows:
\be
|\Phi\ra = \begin{cases}
  \frac{1}{\sqrt{2}}(|1\ra+|2\ra) & \text{for $x<0.115$} \\
\sin\theta_x|0\ra+\frac{\cos\theta_x}{\sqrt{2}}|1\ra+\frac{\cos\theta_x}{\sqrt{2}}|2\ra & \text{for $x>0.115$}
\end{cases}
\ee
\\
where,
\be
x=\frac{\sin2\theta_x-2\sqrt{2}\cos2\theta_x}{36\sin2\theta_x(\sin^2\theta_x-\frac{1}{2}\cos^2\theta_x)}.
\ee

The above relation shows that the nature of the ground state changes at a critical point $x_c=0.115$, which can be a signal of the phase transition. Finally we analyze the case where the symmetry is fully broken and the ground state has the general form as $|\Phi\rangle= \sum_{i=0}^{2} a_i | i\rangle$. In this case, we minimize the energy numerically. Figure \ref{meanfield} demonstrates the mean-field ground state energy per site as a function of 
$x=\frac{2\lambda}{9J}$. As we can see, there is a small kink in the ground state energy curve (a sharp jump in its derivative) at $x_c\approx0.115$. So the results show that the ground state actually respects some permutation symmetries. This jump signals a first order quantum phase transition.  In order to investigate the critical point more accurately, we resort to a more accurate 
approximation technique, i.e. the Perturbative Continuous Unitary Transformation method\cite{wegner,pcut1,pcut2,pcut3} which is the subject of the next subsection.

\subsection{Perturbative Continuous Unitary Transformations}\label{Sec:pcut}
In this section, we briefly review the Perturbative Continuous Unitary Transformation (PCUT) method and apply it to the small- and large-coupling limits of the problem. 

The Continuous Unitary Transformation (CUT) method which was first introduced by Wegner\cite{wegner} in the framework of condensed matter theory, is basically used to 
 diagonalize or block-diagonalize a given Hamiltonian by applying an infinite number of unitary operators to the initial Hamiltonian in a continuous fashion as:
\be \label{H}
H(\ell)=U^{\dagger}(\ell)HU(\ell),
\ee
where $\ell$ is the continuous flow parameter such that $H=H(\ell=0)$ and $H_{\rm eff}=H(\ell=\infty)$ is the (block-) diagonal Hamiltonian. The Hamiltonian is transformed by a unitary operator which its evolution is governed by:
\be \label{u}
\partial_l U(\ell)=-U(l) \eta(\ell).
 \ee
In which $\eta(\ell)$ is the anti-Hermitian generator of the unitary transformation $U(\ell)$. Combining Eq.~(\ref{H}) and (\ref{u}) together one can show that the initial Hamiltonian flows in the form of a differential commutator equation :
\be\label{flow_equation} 
\partial_{\ell} H(\ell)=\comutator{\eta(\ell)}{H(\ell)}.
\ee

The method therefore requires the choice of a suitable generator for the unitary operators to obtain the desirable form of $H_{\rm eff}$. 
Uhrig and Knetter introduced the quasiparticle (QP) conserving generator which is very well suited for our purpose. 
We refer the interested reader to Refs\cite{pcut1, pcut2} for detailed discussions on QP conserving generator. 

The perturbative version of the CUT method (PCUT), can be applied to the Hamiltonians of the form $H=Q + \lambda V $ where the first part of the Hamiltonian, $Q$, is diagonal 
with an equidistant spectrum bounded from below and the second part can be treated as a perturbation ($\lambda$ is the expansion parameter). 
The method further requires that the perturbing part can be written in the form $V =\sum_{n=-N}^{N} T_n$, where $T_n$ increments 
(decrements, if $n<0$) the number of excitations (quasiparticles) by $n$ such that $\comutator{Q}{T_n}=nT_n$ \cite{pcut1}.
Transforming the initial problem by using the QP conserving generator, the effective Hamiltonian is brought to the form that conserves the
number of quasiparticles, $\comutator{H_{\rm eff}}{Q}=0$. The energy spectrum of the system can therefore 
be extracted perturbatively by acting the $H_{\rm eff}$ on the ground state and the multi-particle sectors of the Hilbert space.

\subsubsection{Small-coupling limit ($\lambda \ll J$)}\label{Sec:small-coupling}
In the following we discuss the procedure of applying the PCUT method to the small-coupling limit ($\lambda \ll J$) of the mapped  Kitaev-Potts model, i.e.
the Potts model in transverse magnetic field on the A or B sublattices of Fig.~\ref{mapping}:
\bea
H &=&-J \sum_{i \in A} (\hat{X}_i+\hat{X}^{\dagger}_i) \nonumber \\
&-&\frac{ \lambda} {3} \sum_{ \langle i,j \rangle\in A}  \quad \sum_{r=0} ^{2} \left( (\hat{Z}_i \hat{Z}^{\dagger}_j)^r + (\hat{Z}_j \hat{Z}^{\dagger}_i)^r\right).
\label{maped_model}
\eea
The first term in Eq.~(\ref{maped_model}) is an effective field term which is diagonal. Denoting the local vacuum of each site by $\ket{0}$, the elementary excitations of the 
model for $d=3$ are two separate spin flips labeled by $\ket{1}$ and $\ket{2}$ which correspond to the eigenstates of $X$ operator with $\omega$ and $\omega^{-1}$ eigenvalues, respectively.
Let us note that either of the excitations cost an energy of $3J$. The elementary excitations are energetically indistinguishable. So the first term for $d=3$ has an equidistant spectrum and can 
be regarded as $Q$ for implementation of the PCUT, and the PCUT
results obtained in 1-QP sector of the Hilbert space such as 1-QP gap are degenerate for both of the excitations.

The Potts interaction at the right side of Eq.~(\ref{maped_model}) can be treated as a perturbation $V$ which for $d=3$ is denoted by:
\be
V= -\frac{ 2\lambda} {3} \sum_{ \langle i,j \rangle}  \quad \left( (\hat{Z}_i \hat{Z}^{\dagger}_j) + (\hat{Z}_j \hat{Z}^{\dagger}_i)\right) + C,
\ee
where C is a constant. The perturbing part consists of two-body interactions and can change the number of excitations over the ground state of the effective 
field term $Q$ by $n=\{0,\pm1,\pm2\}$ when it acts on the bonds of the square lattice. Therefore the Hamiltonian (\ref{maped_model}) can be written
as:
\be\label{H_pcut_Tn}
H= Q - x (T_{2}+T_{1}+T_{0}+T_{-1}+T_{-2}),
\ee

where 
\be
Q =\sum_i \frac{ -  (\hat{X}_i+\hat{X}^{\dagger}_i)+2I}{3},
\ee
is the quasiparticle counting operator ($I$ is the identity operator), $x=\frac{2\lambda}{9J}$ is the expansion parameter and $T_n$ operators are given by:

\begin{align}
&T_{+2}=\sum_{ \langle i,j \rangle} |12 \rangle_{i,j} \langle 00| + |21\rangle_{i,j} \langle 00|, \\ \nonumber
&T_{+1}=\sum_{ \langle i,j \rangle}  |22 \rangle_{i,j} \langle 01|+ |22 \rangle_{i,j} \langle 10|\\ \nonumber
 & \qquad +  |11 \rangle_{i,j} \langle 02|+ |11 \rangle_{i,j} \langle 20|, \\ \nonumber
&T_{0}= \sum_{ \langle i,j \rangle} |12 \rangle_{i,j} \langle 21|+ |01 \rangle_{i,j} \langle 10|+ |02 \rangle_{i,j} \langle 20|\\ \nonumber
&\qquad + |21 \rangle_{i,j} \langle 12|+  |10 \rangle_{i,j} \langle 01|+ |20 \rangle_{i,j} \langle 02|.\nonumber
\end{align}
From the hermiticity condition $T_n^\dagger=T_{-n}$, and absorbing the expansion parameter $-x$ in the definition of $T_n$ operators, Eq. (\ref{H_pcut_Tn}) is recast into:
\be 
H= Q + \sum_{n=-2}^{n=2} T_n.
\ee
Under the CUTs, the above Hamiltonian is continuously transformed with the flow parameter $\ell$ as:
\be 
H(\ell)= Q + \sum_{n=-2}^{n=2} T_n(\ell),
\ee
where we wish to reach a situation where $T_n(\ell=\infty)=0$ for all $n\neq0$. In order to fulfil this demand, we choose the quasiparticle conserving generator in 
the following form\cite{pcut1,pcut2}:
\be
\eta(\ell) = T_{+2}(\ell) + T_{+1}(\ell) - T_{-1}(\ell) - T_{-2}(\ell).
\ee
With this choice of generator, the flow Eq.~ (\ref{flow_equation}) can be written as:
{\small
\bea \label{tn}
&\partial_{\ell} & T_0(\ell) = 2[T_{+2}(\ell),T_{-2}(\ell)]+2[T_{+1}(\ell),T_{-1}(\ell)], \\ \nonumber
&\partial_{\ell} & T_{+2}(\ell) = -2 T_{+2}(\ell)+[T_{+2}(\ell),T_{0}(\ell)],\\ \nonumber
&\partial_{\ell} & T_{+1}(\ell) = - T_{+1}(\ell)+2 [T_{+2}(\ell),T_{-1}(\ell)] +[T_{+1}(\ell),T_{0}(\ell)]. \nonumber
\eea}
 Let us stress that the Hamiltonian remains hermition under the unitary transformation. We can therefore calculate $T_{-2}(\ell)$ and $T_{-1}(\ell)$ from
 the hermiticity condition.
Solving the flow equation is still a very cumbersome task because there are an infinite number of terms in $T_n(\ell)$. However, we can tackle the problem by performing a perturbative expansion of the flow equation. We can therefore introduce the expansion of the $T_n$ operators as\cite{pcut3}:
\be
T_n(\ell)=\sum_{i=1}^\infty T_n^{(i)} (\ell),
\ee
where $i$ is the order of perturbation. Using this relation, the perturbative expansion of the flow equation is written as:
{\small
\bea \label{tni} \nonumber
&\partial_{\ell} & T_0^{(k)}(\ell) = 2\sum_{j=1}^{k-1}[T_{+2}^{(j)}(\ell),T_{-2}^{(k-j)}(\ell)]+2\sum_{j=1}^{k-1}[T_{+1}^{(j)}(\ell),T_{-1}^{(k-j)}(\ell)], \\ \nonumber
&\partial_{\ell} & T_{+2}^{(k)}(\ell) = -2 T_{+2}^{(k)}(\ell)+\sum_{j=1}^{k-1}[T_{+2}^{(j)}(\ell),T_{0}^{(k-j)}(\ell)] ,\\ 
&\partial_{\ell} & T_{+1}^{(k)}(\ell) = - T_{+1}^{(k)}(\ell)+2\sum_{j=1}^{k-1} [T_{+2}^{(j)}(\ell),T_{-1}^{(k-j)}(\ell)] \\ \nonumber
&+&\sum_{j=1}^{k-1}[T_{+1}^{(j)}(\ell),T_{0}^{(k-j)}(\ell)].
\eea}
\begin{figure}[t]
\includegraphics[width=7cm,height=6cm]{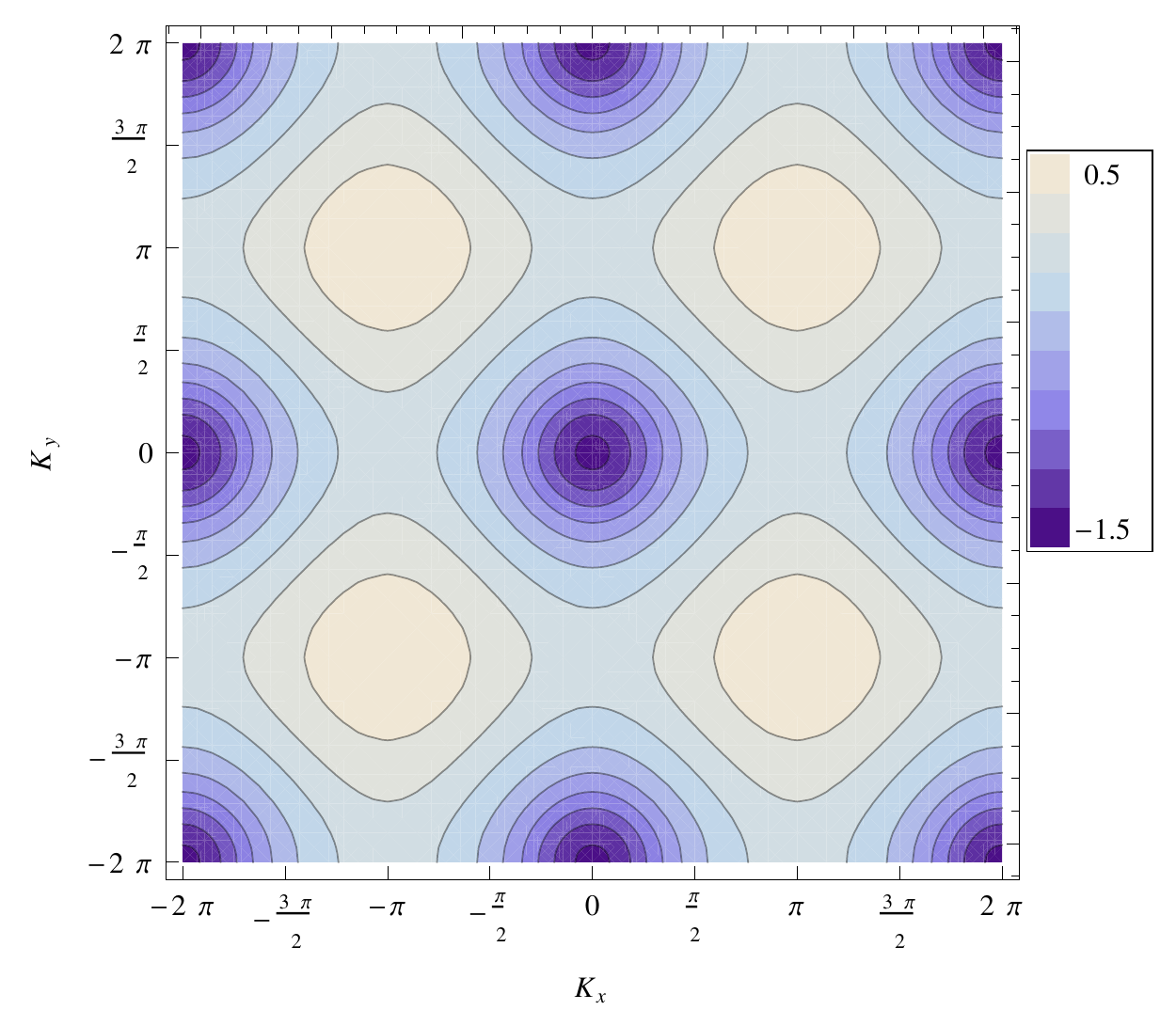}
\caption{(Color Online) 1-QP dispersion of the Kitaev-Potts model in the small coupling regime. The minimum of the dispersion is located at K=(0,0).} \label{dispersion} 
\end{figure}
Solving this set of equations for the initial condition $T_n^{(i)}(\ell=0)=\delta_{1,i}T_n$ and then taking the limit of ($\ell\longrightarrow \infty$), we can obtain $H_{\rm eff}$.
The effective QP conserving Hamiltonian for the small-coupling limit up to order $3$ in perturbation parameter is obtained as:
\bea \nonumber
H_{\rm eff}^{(3)} &=&  Q-x T_{0}+  x^2\comutator{T_{1}}{T_{-1}} + \frac{x^2}{2} \comutator{T_{2}}{T_{-2}}\\ \nonumber
&&-\frac{x^3}{8} \bigg(\comutator{T_2}{\comutator{T_0}{T_{-2}}}+\comutator{\comutator{T_2}{T_0}}{T_{-2}}\bigg) \\
&& -\frac{ x^3}{2}\bigg(\comutator{T_1}{\comutator{T_{1}}{T_{-2}}}+\comutator{\comutator{T_{2}}{T_{-1}}}{T_{-1}}\bigg) \\ \nonumber
&&- \frac{x^3}{2}\bigg(\comutator{T_1}{\comutator{T_0}{T_{-1}}} +\comutator{\comutator{T_1}{T_{0}}}{T_{-1}}\bigg).
\eea
The ground state energy and 1-QP gap can be obtained by acting the effective hamiltonian on 0P and 1P sector of the Q.
We have calculated the ground state energy and 1-QP energy gap of the system in the small-coupling limit up to order 8 in perturbation theory:
\bea
\epsilon^{\rm sc}_{0} &=& -\frac{2}{3}-2 x ^2-x ^3-\frac{17 x ^4}{2}-\frac{847 x ^5}{36}-\frac{18407 x ^6}{144} \\ \nonumber
&&-\frac{15290 \lambda ^7}{27}-\frac{995278817 \lambda ^8}{311040},\\ 
\Delta^{\rm sc} &=& 1-4 x-10 x^2-5 x^3-\frac{1895 x^4}{6}+\frac{14107 x^5}{18}\\ \nonumber
&& -\frac{3572759 x^6}{216}+\frac{26566267 x^7}{324}-\frac{85919559673 x^8}{77760}.
\eea

\begin{figure}[t]
\includegraphics[width=\columnwidth,height=6cm]{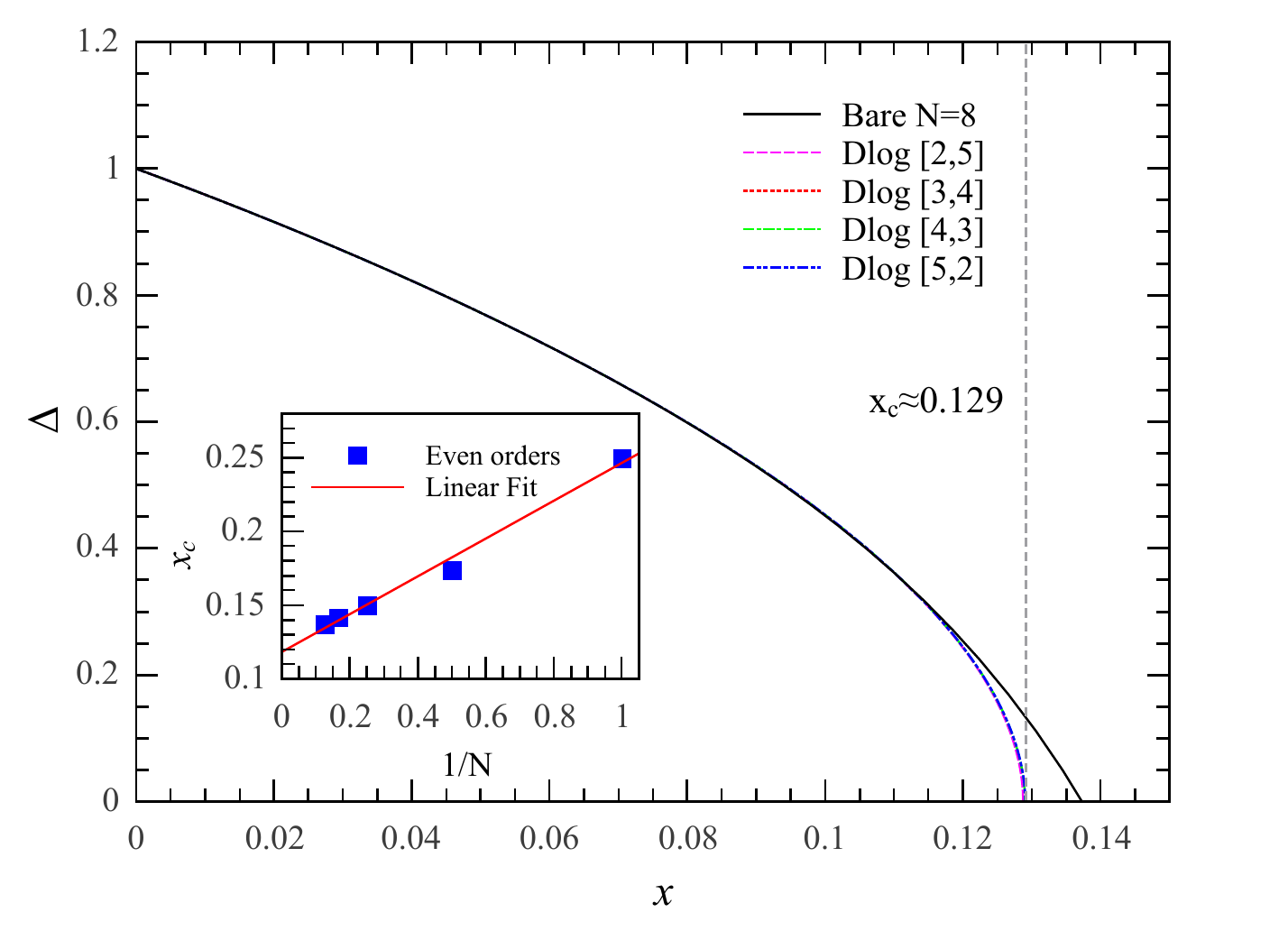}
\caption{(Color Online) The 1-QP gap of the Kitaev-Potts model for different DlogPad\'{e} approximants. 
The critical point highlighted by gray dashed line is consistent with the closure of the gap at $x_c\approx 0.129$. The inset demonstrates
the scaling on the closure of the bare series of the gap for different orders of perturbation.} \label{small-gap} 
\end{figure}
Fig.~\ref{small-gap} illustrates the 1-QP gap of the system in the small-coupling limit as a function of $x=\frac{2\lambda}{9J}$. 
The bare and extrapolated series are 
well converged. Closure of the 1-QP gap occurs at $x_c\approx 0.129$. The inset of Fig.~\ref{small-gap} further demonstrates scaling of the
closure of the bare series of the gap for different orders of perturbation. As one can see, by increasing the order of perturbation the transformation point starts to become smaller until
it converges to $x_c\approx 0.12$ which is fully consistent with the closure of DlogPad extrapolants.

Existence of anyonic excitations in the system is one of the signatures of the topological order. 
As one can see form Fig.~\ref{small-gap}, the anyonc gap is open until in the vicinity of the transition point. It is therefore reasonable to point out that 
the topological order survives during the tuning of the perturbation until the critical point at which the anyons condenses due to the strength of the Potts interaction and 
the topological order breaks down \cite{ssj,ssj2}.

Let us further note that the knowledge of the gap is not solely sufficient to determine the first- or second-order 
nature of the phase transition\cite{ssj,ssj2} and one has to analyze the ground state energy of the system and its derivatives\cite{sachdev} to capture the phase transition 
correctly. We therefore postpone further discussion on the phase transition to Sec.~\ref{QPT} after we calculate the ground state energy series in large-coupling limits.
\begin{figure}[t]
\includegraphics[width=\columnwidth,height=6cm]{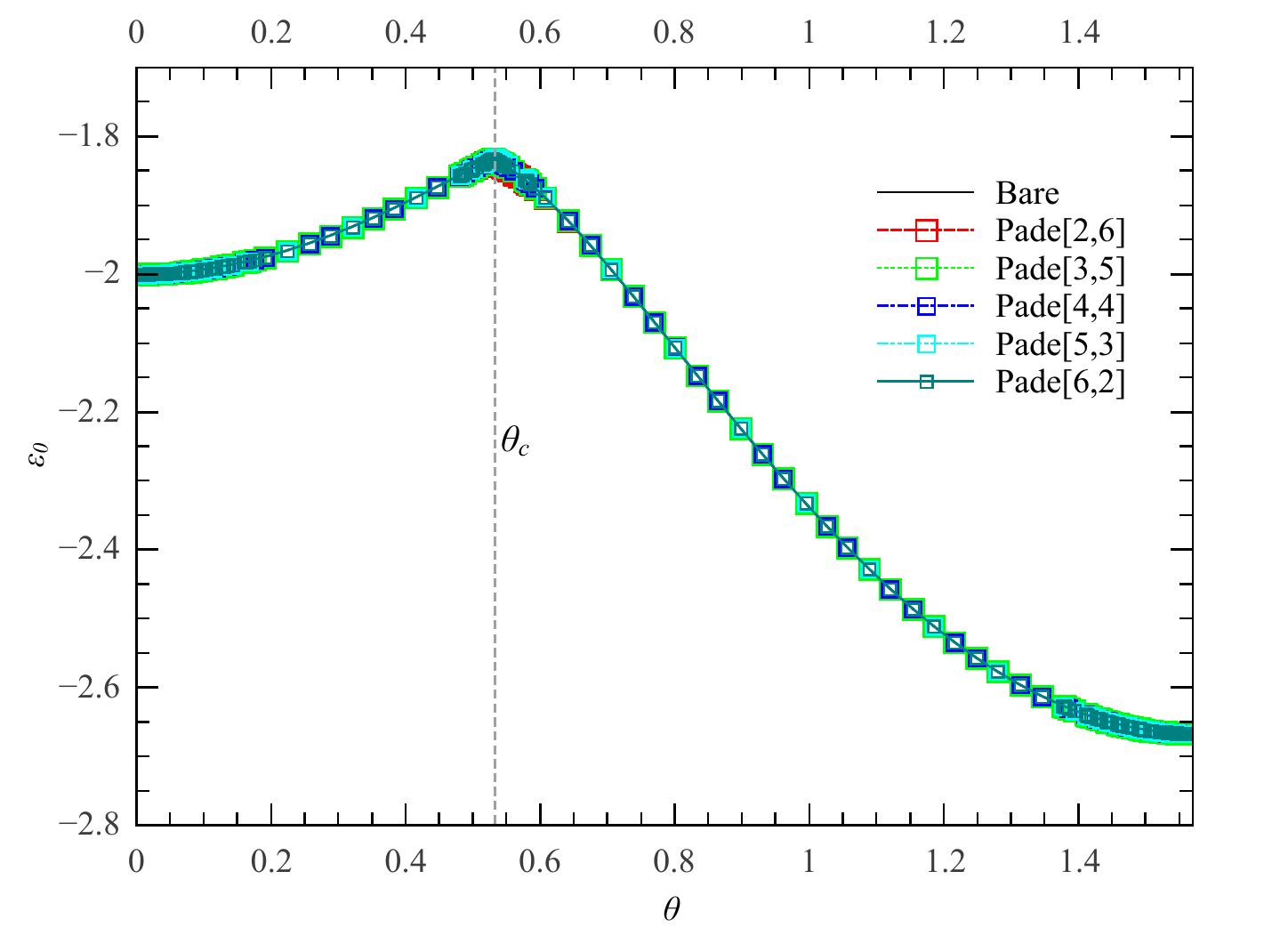}
\caption{(Color Online) The ground state energy per site $\varepsilon_0$ as a function of $\theta$. Solid line corresponds to the bare series of order 8. 
Symbols denote different Pad\'{e} approximants.} \label{gsps} 
\end{figure}
\subsubsection{Large-coupling limit ($\lambda \gg J$)}\label{Sec:larg-coupling}
Following our discussion, we now apply the PCUT method to the large-coupling limit of the problem ($\lambda \gg J$).
For $J=0$,  the Hamiltonian is $3$-State Potts model which is ferromagnetically ordered and has an equidistant spectrum thus satisfying the first 
condition for the PCUT method. The excitations of the model further correspond to the anti-ferromagnetic bonds of the square lattice. 
When $J\neq0$, the effective field term in Eq. (\ref{maped_model}) can be considered as a perturbation which changes the 
number of excitations by $n=\{0,\pm1,\pm2,\pm3,\pm4\}$. 
We can therefore write the Hamiltonian of the model in the large-coupling limit in terms of $T_n$ operators as:
\be
H=Q-h\sum_{n=-4}^{4}T_n,
\ee
where $Q$ is the quasiparticle counting operator defined as:

\be
Q =\frac{2I-\sum_{ \langle i,j \rangle} (\hat{Z}_i \hat{Z}^{\dagger}_j) + (\hat{Z}_j \hat{Z}^{\dagger}_i)}{3},
\ee
and $h=J/2\lambda$ is the expansion parameter. Using the PCUT method, we have calculated the ground state energy per site, $\varepsilon_0$, in the 
large-coupling limit up to order $8$ in perturbation parameter by acting the $H_{\rm eff}$ on the 0-QP sector of the Hilbert space:
\bea
\epsilon^{\rm lc}_{0} &=& -\frac{8}{3}-\frac{h^2}{2}-\frac{h^3}{8}-\frac{19 h^4}{672}-\frac{3 h^5}{128}-\frac{12779 h^6}{846720}\\ \nonumber
&& -\frac{1052987 h^7}{121927680}-\frac{458808396457 h^8}{62768369664000}.
\eea
As we have already mentioned in the previous section, one can obtain a better understanding about the nature of the phase transition by analysis of the 
ground state energy results. We will therefore provide strong evidences for the first-order phase transition in the $\Z_3$ Kitaev-Potts model in the next section.

\begin{figure}[t]
\includegraphics[width=\columnwidth,height=6cm]{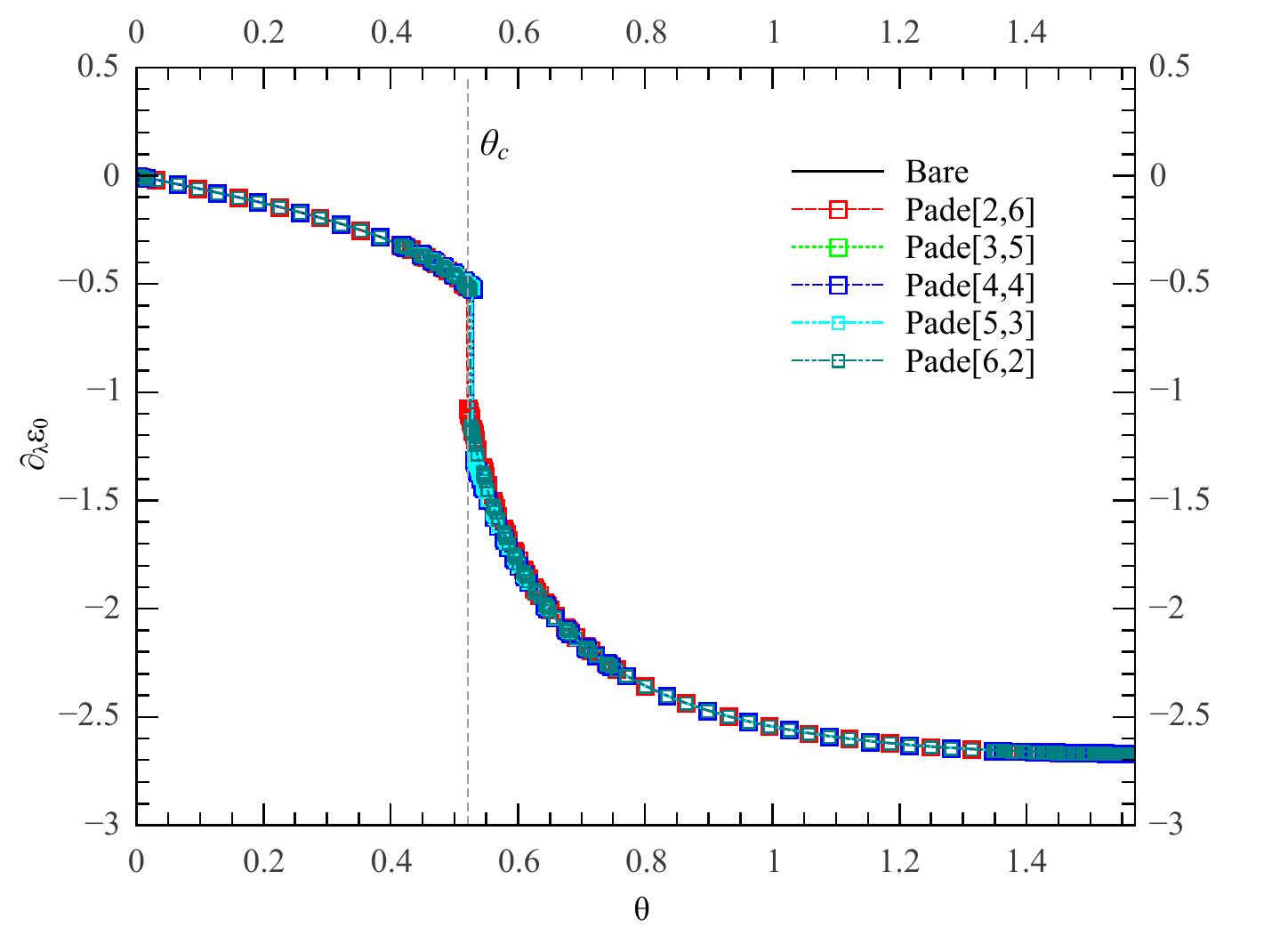}
\caption{(Color Online) First derivative of the ground state energy per site ( $\partial_{\lambda}\varepsilon_0$) as a
function of $\theta$. Solid line corresponds to the bare series of order 8. Symbols denote different Pad\'{e} approximants.} \label{mag} 
\end{figure}

\subsubsection{Analyzing series expansion results}\label{Sec:pcut-results}
\label{QPT}
To investigate the nature of the phase transition, we study the ground state energy per site of the system in the whole range of the expansion parameter by merging the small- and 
large-coupling results. Setting $\lambda=\sin \theta$ and $J=\cos \theta$, we can join the small- and large-coupling results to obtain a complete picture.
Fig.~\ref{gsps} demonstrates the ground state energy per site $\varepsilon_0$ as a function of $\theta$. The small- and large-coupling series cross each other at
$\theta_c\approx0.52$, giving rise to a kink in the $\varepsilon_0$ curve which is fully consistent with the first-order phase transition. 
The location of the kink is essentially the same for different Pad\'{e} extrapolants and has strong agreement with the closure of the 1-QP gap ($x_c=\frac{2}{9} \tan\theta_c\approx 0.129$).

\begin{figure}[t]
\includegraphics[width=\columnwidth,height=6cm]{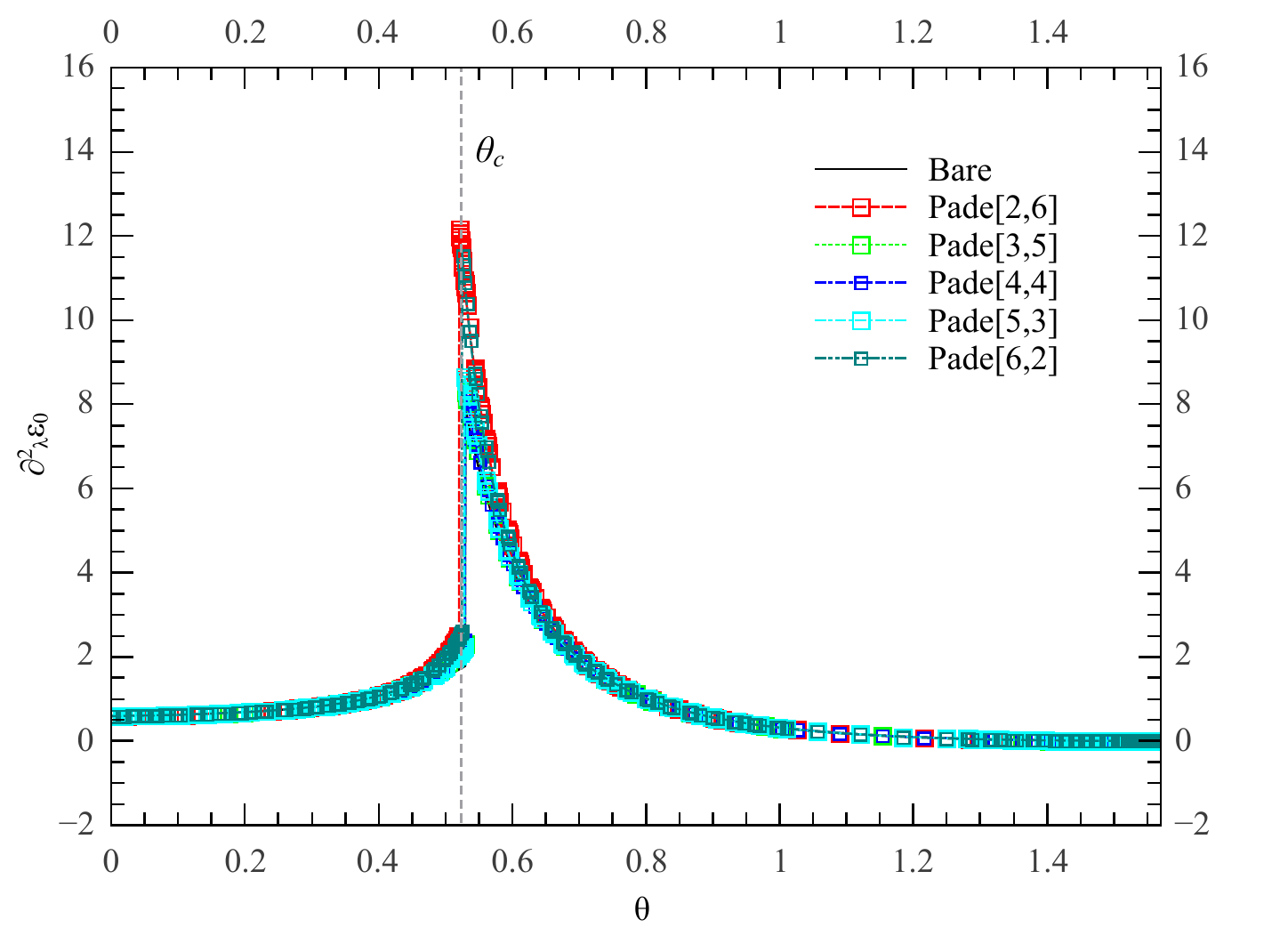}
\caption{(Color Online) Second derivative of the ground state energy per site ( $\partial_{\lambda}^2\varepsilon_0$) as a
function of $\theta$. Solid line corresponds to the bare series of order 8. Symbols denote different Pad\'{e} approximants.} \label{sucep} 
\end{figure}

Using the Feynman-Hellman theorem, we have calculated the first derivative of $\varepsilon_0$ which is equivalent to the magnetization in statistical mechanics. Figs.~\ref{mag} and \ref{sucep}
depict first and second derivatives of $\varepsilon_0$ for different $\theta$ values. The sharp jumps in the first derivative is a clear signature of the first-order 
phase transition \cite{sachdev}. Let us further note that the first order nature of the phase transition in the $\Z_3$ Potts model in transverse magnetic field has already been confirmed using 
series expansion combined with infinite projected entangled-pair state (IPEPS) method in Ref. \cite{orus}. The jump in the second derivative further confirms the location of the transition point which is exactly the same as that of the ground-state
energy for the bare series and different Pade extrapolants.

\subsection{Geometric Measure of Entanglement}\label{Sec:GME}
In this section we calculate a measure of multipartite entanglement, Geometric Measure of Entanglement (GME)\cite{geo}, as another tool for capturing the phase transition.  GME  is a measure of multipartite entanglement in quantum many body systems which measures the distance, Hilbert Schmidt distance, between a given state, $|\Psi(x)\rangle$, and the closest product state, $|P\ra$, as follows:
\be
\rm GME = - \log_2\left( \max_{|P\rangle}|\langle P|\Psi(x)\rangle|^2\right),
\ee
where the maximization is over all product states. This implies that the more entangled the states are, more distance from the set of product states, which is a convex set, they will have. In order to calculate the GME, we have to calculate the ground state of the mapped Hamiltonian, $|\Psi(x)\rangle$, by PCUT procedure up to a specific order of perturbation. The PCUT method transforms the initial Hamiltonian by a unitary transformation ($U(\infty)$) to a block-diagonal form ($H_{\rm eff}$), the basis of the Hilbert space is rotated, such that $H_{\rm eff}$ commutes with $Q$. So they have the same eigenstates. It's also known that  $H_{\rm eff}$'s ground state is the vacuum state of $Q$ with no excitations ($|\tilde{0}\rangle$)\cite{knetter}. The $H_{\rm eff}$  and the initial Hamiltonian, $H$, are unitary equivalent and their ground states are related to each other as follows:
\be
|\Psi(x)\rangle=U(\infty)|\tilde{0}\rangle.
\ee
In order to  determine $U(\ell)$, we expand Eq.~(\ref{u}) as a function of the perturbation parameter $x$:
\be
\partial_l U^{(k)}(\ell)= - \sum_{j=0}^{k-1} U^{(j)}(\ell)  \eta^{(k-j)}(\ell),
\ee
where
\be
\eta^{(k-j)}(\ell)=T_{+2}^{(k-j)}(\ell)+T_{+1}^{(k-j)}(\ell)-T_{-1}^{(k-j)}(\ell)-T_{-2}^{(k-j)}(\ell),
\ee
and $k$ is the order of perturbation. By using Eqs.~(\ref{tni}), and the initial condition $U^{k}(0)= 1 \delta_{k,0}$, one can solve the above equation perturbatively up to second order in $x$. Taking the limit $\ell \rightarrow \infty$, the $U(\infty)$ is found to be:
\begin{align}
U &=1+\frac{x}{2} (T_{+2} -T_{-2}) +x (T_{+1}-T_{-1}) \nonumber \\
&+\frac{x^2}{2}(T_{+1}-T_{-1})^2+\frac{x^2}{8}(T_{+2}-T_{-2})^2 \nonumber \\
&+\frac{x^2}{4}[T_0,(T_{+2}+T_{-2})]+x^2[T_0,(T_{+1}+T_{-1})] \nonumber \\
&+\frac{x^2}{6} (T_{+1}T_{+2}+T_{-1}T_{-2} +2T_{+2}T_{+1}+2T_{-2}T_{-1})\nonumber \\
&+\frac{x^2}{2}([T_1,T_{-2}]+[T_{-1},T_2]-T_{-2}T_1-T_2T_{-1}).
\end{align}

The ground state of the initial Hamiltonian $H$ is therefore given by:
\begin{align}
|\Psi(x)\rangle &=U|\tilde{0}\rangle=(1-\frac{4n x^2}{8})|\tilde{0}\rangle + \frac{x}{2} \sum_{4n} \big| 1\quad 2\big\rangle\\ \nonumber
&+\frac{x^2}{4} \Bigg(\sum_{4n} \big| 1\quad 2\big\rangle+ 2\sum_{8n} \Big| \begin{array}{cc}
1 & 0  \\
0 & 2 
\end{array} \Big\rangle +2 \sum_{4n} \big| 1 \quad 0 \quad 2 \big\rangle  \Bigg) \\  \nonumber
&+ \frac{x^2}{6} \Bigg( 2 \sum_{4n} \Big| \begin{array}{cc}
2 & 2  \\
2 & 0 
\end{array} \Big\rangle +2\sum_{4n} \Big| \begin{array}{cc}
1 & 1  \\
1 & 0
\end{array} \Big\rangle  \\ \nonumber
&+ 2 \sum_{2n} \big| 2 \quad 2 \quad 2 \big\rangle +2 \sum_{2n} \big| 1 \quad 1 \quad 1 \big\rangle \Bigg) \\ \nonumber
&+ \frac{x^2}{ 8}\Bigg( \sum_{8n(2n-7)} \Big| \begin{array}{cc} 
1 & 2 \\
1 & 2 
\end{array} \Big\rangle \Bigg).
\end{align}
This state is a superposition of some product states, these product states are eigenstates of $Q$. The first term, $|\tilde{0}\rangle$ refers to the ground state of $Q$ with no excitations, which is the state of all spins in the  $\hat{X}$ eigenstate with $+1$ eigenvalue, $|0\ra$.
$\big| 1\quad 2\big\rangle$ is the state of all the spins in $|0\ra$ state, except for the two of them, which are nearest neighbour, and have the states of $|1\ra$ and $|2\ra$. The summation must be done over all the states of this kind, and one should note that in a lattice with $n$ vertices the number of this kind of states are 2 times the number of bonds, i.e. $2\times2n=4n$. Furthermore, $\Big| \begin{array}{cc}
1 & 0  \\
0 & 2 
\end{array} \Big\rangle$ and $\big| 1 \quad 0 \quad 2 \big\rangle$ denote the state with all spins in the $|0\ra$ state, except for the two of them, which are next nearest neighbours and have the states of $|1\ra$ and $|2\ra$. The last term, $\Big| \begin{array}{cc} 
1 & 2 \\
1 & 2 
\end{array} \Big\rangle$, also refers to the state with all spins in $|0\ra$ state except for the four of them. The state of these four spins consist of two clusters of $|1 \quad 2\ra$, where these two clusters can be separate or not.
\begin{figure}[t]
\begin{center}
\includegraphics[width=\columnwidth]{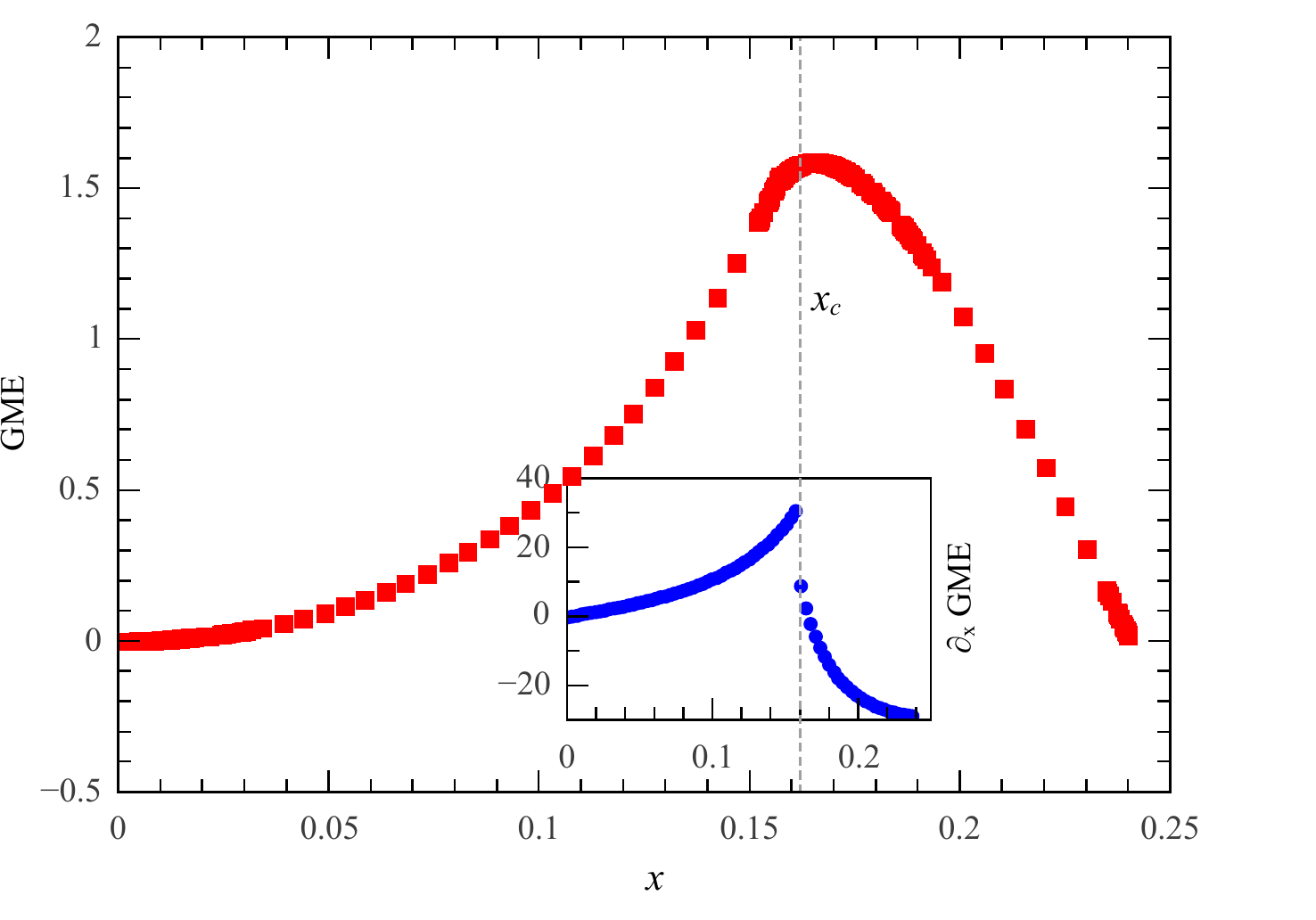}
\caption{(Color Online)GME as a function of $x$ for n=25. Convexity of GME changes close to the $x_c\approx0.16$. The inset further demonstrates the derivative of GME as a function of $x$. The sharp jump in the figure is a clear signature of phase transition.}\label{geo}
\end{center}
\end{figure} 

$|\Psi(x)\rangle$ is exact up to the second order in perturbation parameter $x$ and is normalized. We have also calculated the ground state of the initial Hamiltonian 
up to the fourth order, numerically and calculated the GME. Since the ground state has translational symmetry, the closest product state also preserves 
this symmetry\cite{symmetry}. We can therefore perform the maximization only over the states in the form $|P \rangle =|\phi\rangle^{\otimes n}$, where $|\phi\rangle  =\cos(\theta)|0\rangle + e^{-i \alpha} \sin(\theta)\sin(\varphi)|1\rangle+e^{-i\beta} \sin(\theta) \cos(\varphi)|2\rangle$. In other word, the maximization is only over 4 parameters which makes the numerical calculation of the GME possible. Fig.~\ref{geo} illustrates the geometric measure of entanglement as a function of $x$. 
As we can see, the convexity of the GME changes sign from positive to negative close to a critical point, $x_c \approx 0.16$. 
The sharp jump in the derivative of GME captures the phase transition more clearly, since it's known that for two-dimensional systems discontinuity in 
the derivative of multipartite entanglement leads to a quantum phase transition\cite{evi}. The critical point is very close to the one obtained from 
analysis of the ground state energy and the gap i.e. $x_c\approx0.129$. The slight difference is likely the consequence of the difference in orders of perturbation. 
This can be best inferred by reminding the scaling behavior of the closure of 1-QP gap (inset of Fig.~\ref{small-gap}) which predicts that increasing the order of 
perturbation makes the result to become smaller and converge to $\approx0.12$. The same fact should also holds here for the location of sharp jump in the 
derivative of the GME.

\section{Conclusion}\label{Sec:conclude}
\label{conclud}
The $\Z_d$ Kitaev model is a system with a topologically ordered ground state which, when complemented with some other resources like magic state distillation \cite{bravyi,earl} or measurements \cite{pachas},  is well suited for the purpose of universal quantum 
computation without resorting to non-Abelian groups. It is therefore of great interest to study the stability and robustness of the topological 
phase of the model in the presence of external perturbations.
In this paper, we studied the competition between the topological order induced by the $\Z_3$ Kitaev model and the local order induced by the 3-State Potts model on
the square lattice. 
We showed that the Kitaev model in the presence of the Potts interaction is mapped to Potts model in a transverse magnetic field. 
Using the  high-order series expansion based on the continuous unitary transformations in the small- and large- Potts couplings, we showed that 
the topological phase breaks down to a non-topological phase with local order parameter through a first-order quantum phase transition at $x_c=\frac{2\lambda_c}{9J_c}\approx0.129$.
Our results were further in good agreement with the mean-field approximation results at $x_c\approx0.115$. Computing the Geometric Measure of Entanglement also shows that the derivative of GME has a sharp jump very close to the critical point and also, the convexity of GME  changes sign from positive to negative in this point.
\\
\section{Acknowledgements}
R.M. acknowledges R. Haghshenas for fruitful discussions.

{}
\end{document}